\documentclass{aa}
\usepackage{graphicx}
\usepackage{natbib}
\usepackage[varg]{txfonts}
\newcommand{\hi}{{\sc H\,i}}
\newcommand{\atlas}{ATLAS$^{\rm 3D}$}

\begin{document}

\title{The H{\,\Large I} Tully-Fisher Relation of Early-Type Galaxies}
\titlerunning{The H{\,\scriptsize I} Tully-Fisher relation of Early-Type Galaxies}
\authorrunning{M. den Heijer et al.}
\author{Milan den Heijer\inst{1,2}\thanks{Member of the International Max Planck Research School (IMPRS) for
Astronomy and Astrophysics at the Universities of Bonn and Cologne}, 
Tom A. Oosterloo\inst{3,4}\thanks{Corresponding author: oosterloo@astron.nl}, 
Paolo Serra\inst{5},  
Gyula I.G. J\'{o}zsa\inst{6,7,1},
J\"{u}rgen Kerp\inst{1},\\
Raffaella Morganti\inst{3,4}, 
Michele Cappellari\inst{8},
Timothy A. Davis\inst{9},
Pierre-Alain Duc\inst{10},\\
Eric Emsellem\inst{11}, 
Davor Krajnovi\'{c}\inst{12},
Richard M. McDermid\inst{13,14},
Torsten Naab\inst{15},\\
Anne-Marie Weijmans\inst{16} and 
P. Tim de Zeeuw\inst{11,17}
}

\institute{
Argelander Institut f\"ur Astronomie (AIfA), University of Bonn, Auf dem H\"ugel 71, 53121 Bonn, Germany
\and
Max Planck Institut f\"ur Radioastronomie (MPIfR), Auf dem H\"ugel 69, 53121 Bonn, Germany
\and
ASTRON, the Netherlands Institute for Radio Astronomy, Postbus 2, 7990 AA Dwingeloo, The Netherlands
\and
Kapteyn Astronomical Institute, University of Groningen,  Postbus 800, 9700 AV Groningen, The Netherlands
\and
CSIRO Astronomy and Space Science, Australia Telescope National Facility, PO Box 76, Epping, NSW 1710, Australia
\and
SKA South Africa Radio Astronomy Research Group, 3rd Floor, The Park Park Road Pinelands, 7405, South Africa
\and
Rhodes University, Department of Physics and Electronics, Rhodes Centre for Radio Astronomy, Techniques \& Technologies, PO Box 94, Grahamstown 6140, South Africa
\and
Sub-Department of Astrophysics, Department of Physics, University of Oxford, Denys Wilkinson Building, Keble Road, Oxford OX1 3RH, UK
\and
Centre for Astrophysics Research, University of Hertfordshire, Hatfield, Herts AL1 9AB, UK
\and
Laboratoire AIM Paris-Saclay, CEA/IRFU/SAp - CNRS - Universit\'{e} Paris Diderot, 91191 Gif-sur-Yvette Cedex, France
\and
European Southern Observatory, Karl-Schwarzschild-Stra\ss e  2, 85738 Garching, Germany
\and
Leibniz-Institut f\"ur Astrophysik Potsdam (AIP), An der Sternwarte 16, 14482 Potsdam, Germany
\and
Department of Physics and Astronomy, Macquarie University, Sydney, NSW 2109, Australia
\and
Australian Gemini Office, Australian Astronomical Observatory, PO Box 915, Sydney, NSW 1670, Australia
\and
Max-Planck-Institut f\"{u}r Astrophysik, Karl-Schwarzschild-Stra\ss e 1, 85741 Garching, Germany
\and
School of Physics and Astronomy, University of St Andrews, North Haugh, St Andrews KY16 9SS, UK
\and
Sterrewacht Leiden, Leiden University, Postbus 9513, 2300 RA Leiden, the Netherlands
}

\date{Received 3 July 2015 / Accepted 14 July 2015}

\abstract{
We study the \hi\ $K$-band Tully-Fisher relation and the baryonic Tully-Fisher relation for a sample of 16 early-type galaxies, taken from the \atlas\ sample, which all have very  regular \hi\ disks extending  well beyond the optical body ($\gtrsim 5\,R_{\rm eff}$). We use the kinematics of these  disks to estimate the circular velocity at large radii for these galaxies. We find that the Tully-Fisher relation for our early-type galaxies is offset by about 0.5-0.7 mag from the relation for spiral galaxies, in the sense that early-type galaxies are dimmer for a given circular velocity.  The residuals with respect to the spiral Tully-Fisher relation  correlate with estimates of the stellar mass-to-light ratio, suggesting that the offset between the relations is mainly driven by differences in stellar populations.\\
We also observe a small offset between our Tully-Fisher relation  with the relation derived for the \atlas\ sample based on CO data representing the galaxies' inner regions ($\lesssim\,$1 $R_{\rm eff}$). This indicates that the circular velocities at large radii  are systematically 10\% lower than those near 0.5-1 $R_{\rm eff}$, in line with recent determinations of the shape of the mass profile of early-type galaxies.   \\
The baryonic Tully-Fisher relation of our sample is distinctly tighter than the standard one,  in particular when using  mass-to-light ratios based on dynamical models of the stellar kinematics.  We find that the early-type galaxies fall on the spiral baryonic Tully-Fisher relation if one assumes ${M}/L_{K}=0.54$ $M_\odot/L_\odot$ for the stellar populations of the spirals, a value similar to that found by recent studies of the dynamics of spiral galaxies.  Such a mass-to-light ratio for spiral galaxies would imply that their disks  are 60-70\% of maximal. Our analysis increases the range of galaxy morphologies for which the baryonic Tully-Fisher relations holds, strengthening previous claims that it is a more fundamental scaling relation than the classical Tully-Fisher relation.
} 


\maketitle

\section{Introduction}

The Tully-Fisher relation (hereafter, TFR;  \citealt{tully1977}) is an established fundamental scaling relation for spiral galaxies. It was originally discovered as a tight correlation between the width of the integrated \hi\ spectrum and the absolute magnitude, but is primarily an empirical relation between  circular velocity and luminosity. Explaining the TFR is fundamental for understanding how galaxies form and evolve. Moreover, since luminosity is distance dependent, while circular velocity is not, a historically important application of the TFR is the determination of distances in the local Universe. 

Since its discovery, there has been significant debate about how to interpret the TFR in terms of more fundamental galaxy properties such as total stellar mass. For this reason, modern TFR studies tend to observe galaxies at infrared wavelengths, where luminosity is a better tracer of stellar mass. An even more fundamental relation might exist between the circular velocity and total baryonic mass, the baryonic TFR (\citealt{walker1999, mcgaugh2000}). For example, while the TFR becomes less-well defined  for smaller galaxies (circular velocities below 50 km s$^{-1}$), \citet{mcgaugh2012} shows that these systems follow the same baryonic TFR as brighter ones. 

The study of the TFR as a function of galaxy morphological type can clarify to what extent the slope, intercept and scatter of the relation are determined by stellar $M/L$ or shape of the mass distribution. Earlier studies indicate that early-type spirals are fainter for a given circular velocity than late-type spirals (\citealt{roberts1978,russell2004, 2009ApJ...705.1496S}). However, this difference reduces when stellar population effects are minimised by considering near-infrared instead of optical photometry (e.g., \citealt{aaronson1983}; \citealt{noordermeer2007}). One other important finding is that the shape of the TFR, in particular whether the TFR exhibits a 'kink' towards high circular velocities for more luminous galaxies,  depends on how and at which radius the circular velocities are measured, reflecting systematic differences in the shape of the mass distribution and hence of the rotation curve \citep{noordermeer2007,williams2010}.

Recently, TFR studies have been extended to early-type galaxies (E and S0, hereafter ETGs). These objects are important because by comparing their TFR to that of spirals we have a larger 'baseline' for investigating  how the stellar mass-to-light ratio and the shape of the total mass distribution influence the TFR. \citet{2006MNRAS.373.1125B} studied the $B$- and $K$-band TFR of lenticular (S0) galaxies and  found that in both bands the TFR for S0's lies below the TFR for spiral galaxies (1.2 mag in $K$-band). Hence, S0 galaxies rotate faster and/or are dimmer. \citet{williams2010} also found an offset based on a study of 14 spirals and 14 S0s using various  velocity tracers and estimate it to be 0.53 mag in the $K_S$-band. \citet{Cortesi2013} performed a study of the S0 TFR using stellar kinematics from planetary nebulae spectra and found the S0 TFR to be offset from the spiral TFR by a full magnitude in $K$-band. Similarly, \citet{rawle2013} find an offset in the $K_{\rm s}$ band between the TFR of spirals and ETGs in the Coma cluster of about 0.8 mag. Recently, \citet{2014MNRAS.440.3491J} used the ionised gas emission of 24 ETGs to construct a TFR and found that their ETGs are 1.7 mag fainter than spirals in the $B$-band.

Until recently,  studies of the ETG TFR have been challenging because such galaxies are typically gas-poor compared to spirals and measuring their circular velocity is less straightforward, in particular at large radii. However, recent surveys, notably the SAURON \citep{dezeeuw2002} and the \atlas\ projects \citep{cappellari2011}, have shown that a significant fraction of  ETGs host significant amounts of molecular and atomic neutral gas, often in the form of a regularly rotating disk or ring \citep{morganti2006,oosterloo2010,young2011,serra2012,davis2013,alatalo2013}. As part of the \atlas\ project, \citet{davis2011} studied the CO TFR and their most important findings are that in many ETGs CO can be used as a tracer for the circular velocity of the flat part of the rotation curve and that the CO TFR is offset from the TFR for spirals by about 1 mag in $K$-band. The CO disks used by \citet{davis2011} typically extend to about 1 $R_{\rm eff}$ and, therefore, the TFR in their work reflects the situation for the inner regions of ETGs. As has been found by, e.g., \citet{noordermeer2007},  the shape of the observed TFR depends on at which radius the circular velocities are measured, reflecting the fact that the shape of  rotation curves varies systematically with galaxy mass and morphology. Here we take advantage of the large size of many of the \hi\ discs found by \citet{morganti2006}, \citet{oosterloo2010} and \citet{serra2012} around ETGs, and study the TFR of these galaxies using circular velocities measured at very large radius ($\gtrsim 5R_{\rm eff}$). 

This paper is structured as follows. In Section 2 we describe the sample and \hi\ data. In Section 3 we discuss the methods used to estimate the \hi\ circular velocity. In Section 4 we present the \hi\ $K$-band TFR and a comparison with the CO TFR from \citet{davis2011} and with TFRs for spiral galaxies. In Section 5 we present the \hi\ baryonic TFR and in Section 6 we summarise our findings and present the conclusions.

\section{Sample and data}
\label{sample}

This work is based on the data from the \atlas\ \hi\ survey of ETGs carried out with the Westerbork Synthesis Radio Telescope (WSRT) by \citet{serra2012}. This survey targeted a volume-limited sample of 166 morphologically selected ETGs with distance $D<42$ Mpc and brighter than $M_K=-21.5$. This sample consists of those galaxies of the larger, volume-limited \atlas\ sample which are observable with the WSRT and includes the  \hi\ data of \citet{morganti2006} and \citet{oosterloo2010}. We refer to \citet{cappellari2011} and \citet{serra2012} for more details on the sample selection and on the \hi\ observations. 

\citet{serra2012}  report the detection of \hi\ in 53 ETGs. In particular, in 34 objects the \hi\ is found to form a rotating disc or ring with radii varying from a few to tens of kpc (classes $d$ and $D$ in that paper). These systems exhibit relatively regular \hi\ kinematics (see also \citealt{serra2014} for a discussion of the kinematics of these \hi\ disks) and are therefore the starting point for our TFR study as these \hi\ disks offer the possibility to derive accurate rotation velocities. A second step in the data selection has been to perform a harmonic decomposition of the velocity field of the outer regions of these 34 galaxies using the \texttt{KINEMETRY} software of \citet{krajnovic2006}. The velocity fields used are those published by \citet{serra2014}.  A harmonic decomposition of the velocity field gives information about to what extent   circular rotation dominates the observed kinematics  compared to other, non-circular motions. Our selection criterion was that the amplitude of the non-circular motions must be less than 10\% of the circular rotation. This eliminates cases where the \hi\ in the outer regions is not sufficiently settled in a disk, or galaxies where strong streaming motions due to a bar might affect the results.
Moreover, since we aim to study the TFR using velocities obtained at large radius, we exclude  galaxies in which the \hi\ component is not very extended compared to the optical size.

These criteria reduce the sample of 34 objects by excluding galaxies with signs of interaction (NGC~3619, NGC~5103, NGC~5173), galaxies with complex \hi\ kinematics (e.g., strong warps or non-circular motions; NGC~2594, NGC~2764, NGC~4036, NGC~5631, UGC~03960, UGC~09519) and galaxies whose \hi\ disc is smaller than or comparable to the stellar body (NGC~3032, NGC~3182, NGC~3414, NGC~3489, NGC~3499, NGC~4150, NGC~4710, NGC~5422, NGC~5866, UGC~05408). This leaves us with 15 galaxies. We add to this sample NGC~2974, which is part of the \atlas\ sample but not of the \hi\ survey  of \citet{serra2012} because of its low declination.

\begin{figure}
\includegraphics[width=8.3cm]{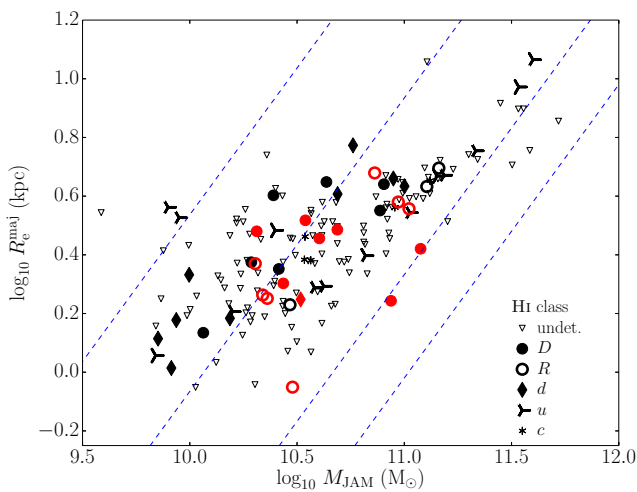}
\caption{Mass-size plane for the galaxies from the sample in this paper (the red symbols) compared to the full sample from \citet{serra2012}. Data are taken from Cappellari et al.\ (2013b) while the definition of the \hi\ classes is very similar to the one used by \citet{serra2012}: $D$: large \hi\ disk (larger than optical body), $R$ large \hi\ ring (larger than optical body), $d$: small \hi\ disk (confined within optical body), $u$: unsettled (tails, streams), $c$: scattered \hi\ clouds. The blue, dashed lines give loci of constant velocity dispersion. }
\label{mass-size}
\end{figure}

Since we consider only relatively \hi-rich galaxies with regular kinematics, we investigate whether this introduces any bias.  In Fig.\ \ref{mass-size} we show the mass-size relation  of the galaxies from \citet{serra2012}  with the sub-sample selected by us for this paper indicated (the data are taken from \citealt{cappellari2013b}).  One can see that very massive and very small galaxies appear to be missing from our sub-sample. The main consequence is that  the range of circular velocities over which we can study the TFR is somewhat limited. Furthermore, Cappellari et al.\ (2013a) showed that the bulge-to-disc ratio, which affects the shape of the rotation curve, changes systematically with velocity dispersion (represented by the blue, dashed lines in Fig.\ 1). The figure shows that our TFR sample nearly covers the entire range of velocity dispersion of the full sample. Therefore, we conclude that our selection has not introduced any major bias. 

In order to construct a TFR, we need the \hi\ circular velocities for these 16 galaxies. For NGC~2685 and NGC~2974 we use the results of \citet{jozsa2009} and \citet{weijmans2008}, respectively.  For the remaining 14 galaxies we use the methods described in Section 3. Since the finite spatial resolution of our observations gives rise to beam-smearing in the inner regions of the velocity fields,  we do not derive complete rotation curves for the galaxies. Instead, we focus on measuring the circular velocity at large radius since this is the relevant quantity for our TFR.  

For the sample galaxies we adopt the absolute magnitudes and distances listed in \citet{cappellari2011}. We refer to that paper for details on the photometry and on the distance estimate of all galaxies. We assume that the main uncertainties in the $K$-band magnitudes, and of the stellar masses we derive from these magnitudes, are caused by distance uncertainties. We  estimate  these from the average deviation of redshift distances from the NED-D redshift-independent distances\footnote{http://nedwww.ipac.caltech.edu/Library/Distances/} which we find to be approximately 21 percent. This implies an  uncertainty in the absolute magnitudes of 0.41 mag.

\begin{figure*}
\hfill
\includegraphics[width=84mm]{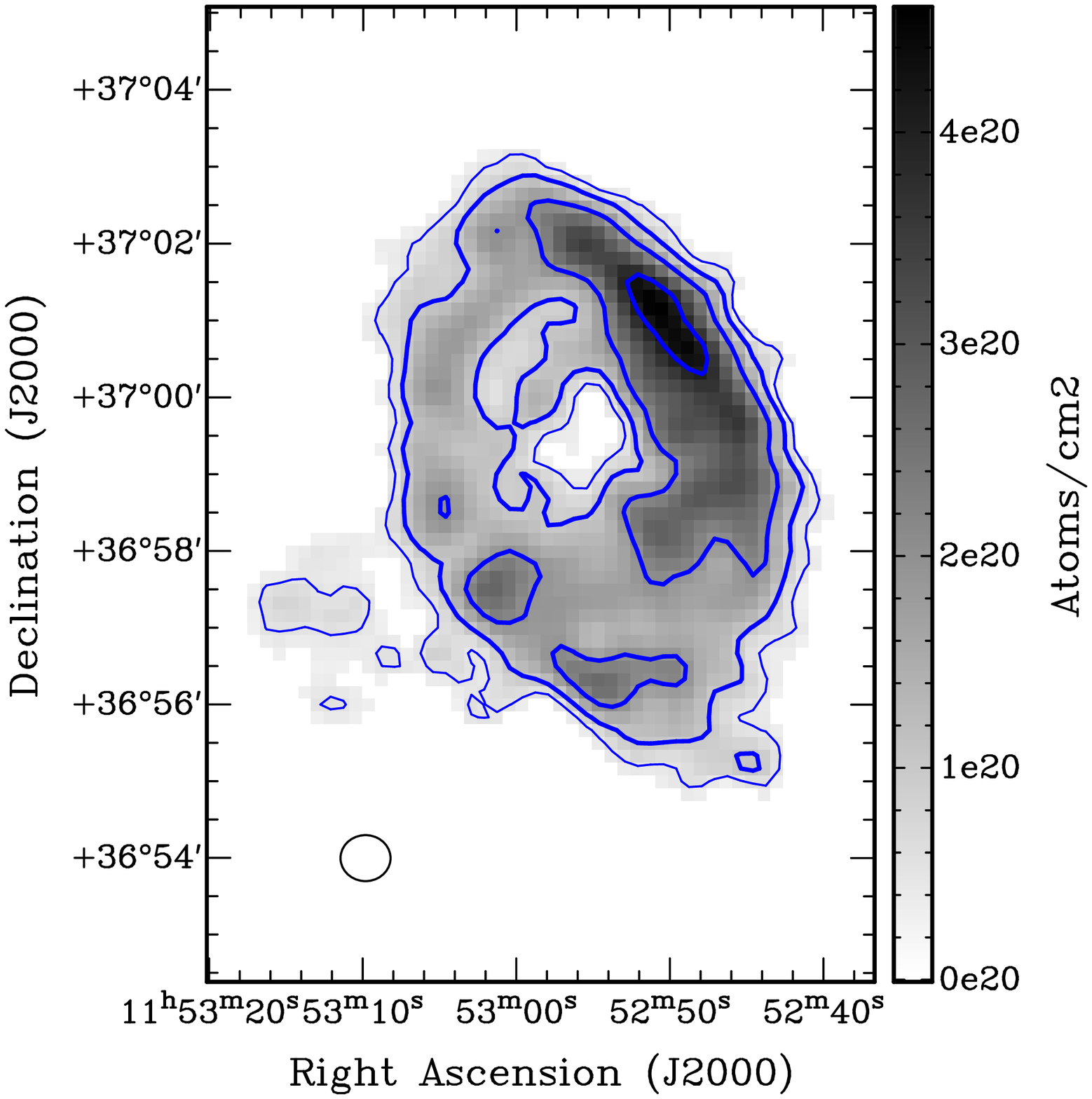}
\hfill
\includegraphics[width=84mm]{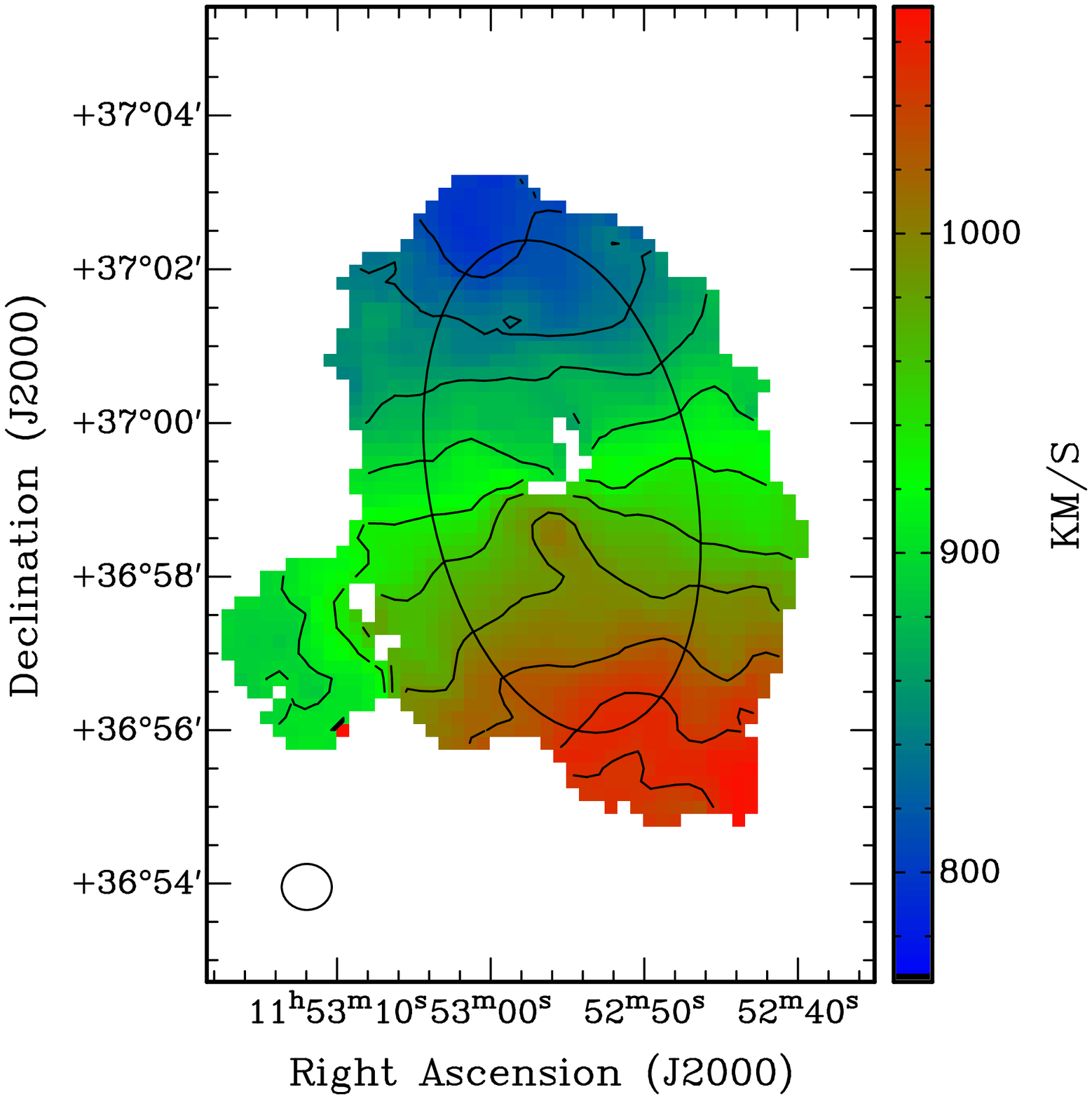}
\hfill
\caption{\textit{Left}: NGC 3941 column density map. The contours are $0.5,~1,~2,~4 \times 10^{20}~\mathrm{cm}^{-2}$. The ellipse on the bottom-left describes the the beam shape. \textit{Right}: First-moment velocity field. The contours range from 780 to 1080 km/s with 30 km/s steps. The overlaid ellipse describes the model geometry (kinematic centre, position angle, inclination and radius). The ellipse on the bottom-left describes the the beam shape.}
\label{n3941nhi}
\end{figure*}

\section{Methods}

An important aspect of the TFR is what quantity should be used as velocity. Historically, the TFR has been constructed using the width of the integrated \hi\ spectrum. However, more recent studies tend to use directly the circular velocity, which can be estimated using a variety of methods including \hi, CO and H$\alpha$ resolved rotation curves (e.g., \citealt{verheijen2001,davis2011,mcgaugh2001}), planetary nebulae \citep{Cortesi2013} or dynamical modelling \citep{williams2009}. These different methods may result in estimates of the circular velocity at different points along the rotation curve, and it is important to check that independent methods deliver consistent results. Recent analyses have indeed highlighted the importance of the rotation curve shape for TFR studies \citep{noordermeer2007}. We reiterate that in this paper we focus on circular velocities for radii much larger than the optical  size of the galaxies.
 
To estimate the \hi\ circular velocity, we start by analysing the moment-1 velocity fields produced by the WSRT data reduction pipeline described in \citet{serra2012} and that are shown in \citet{serra2014}. However, these velocity fields do not always represent the \hi\ kinematics accurately, in particular for objects with a projected size small enough to suffer from beam-smearing or for galaxies where the outer disk is significantly warped. For these objects we attempt to take beam-smearing effects into account by analysing the \hi\ data cube directly. Which technique we have used for which galaxy is indicated in Table \ref{tfr_data}.  We discuss the analysis  procedure applied to velocity fields and to \hi\ cubes in Sections \ref{velfield} and \ref{datacubes}, respectively. Table \ref{tfr_data} lists the resulting circular velocity of all galaxies in our sample. As mentioned in Section \ref{sample}, given the size of the \hi\ beam relative to the extent of the galaxies, we do not attempt to derive complete rotation curves covering all radii. To construct the TFR, we  focus on deriving a single circular velocity for the outermost point of the rotation curve. The radius at which we were able to determine this velocity differs from galaxy to galaxy and ranges from 8 to 28 kpc (15 kpc on average) or $R/R_{\mathrm{eff}}$ from 3.4 to 13.7 (7.3 on average).

\subsection{Velocity field analysis}
\label{velfield}

For galaxies for which the \hi\ is well resolved, we derive the \hi\ circular velocity at large radius by fitting tilted-ring models to the velocity field (\citealt{begeman1987}) using the \texttt{rotcur} task in the \texttt{GIPSY} package (\citealt{vanderhulst1992}). 
This task performs a least-square fit to the velocity field by modelling the galaxy as a set of rings with increasing radius while varying the following parameters of each ring: centre, systemic velocity, inclination, position angle and circular velocity. We adopt a spacing between the rings equal to the \hi\ beam size. As starting parameters for the fitting, the inclination is based on the axis ratio of the \hi\ column density map, while the initial position angle, circular- and systemic velocities are estimated by using position-velocity diagrams. In the fitting we used the following strategy. Firstly, we vary only the coordinates of the centre for all rings together and then fix them for the rest of the fitting procedure. Secondly, we fit all  parameters and fix the systemic velocity of  subsequent fits to one value for all rings based on this fit. Finally, we solve for the inclination, position angle and circular velocity ($v_{\mathrm{circ}}$) for the outer rings.

The uncertainty in the circular velocity is largely determined by the error in  the inclination. It is therefore important to estimate the latter. To be able to assess the accuracy of the best-fitting parameters we follow a two-step strategy. First, we fit, for each ring,  higher order harmonic terms to the velocity field:
\begin{equation}
v(R,\theta) = v_{\mathrm{sys}} + \sum_{k} c_k(R) \cos{k \theta} + s_k(R)  \sin{k \theta},
\label{harmonicseries}
\end{equation}
where $v_{\mathrm{vsys}}$ is the systemic velocity, $c_k(R)$ and $s_k(R)$ are the harmonic amplitudes of order $k$ and $\theta$ is the azimuthal angle in the plane of the galaxy and $v_{\rm circ} = c_1/\sin i$. We use the first four orders of this expansion. This fit is performed using the routine \texttt{reswri} from the \texttt{GIPSY} package \citep{schoenmakers1997}. Following \citet{franx1994}, we use the fact that an error in the inclination angle can be directly related to the $c_3$-term. 
We make use of this  by  estimating the error in the inclination by constructing velocity fields for the outer rings for a range of inclinations  around the value obtained from the tilted-ring fit  and see for which inclinations we detect a non-zero $c_3$ term in the difference between the observed and model velocity field.

Summarising our method, we first fit a regular tilted-ring model to estimate the rings' centre, systemic velocity, position angle and inclination as described above. We then use these values as input for the harmonic decomposition, which we run multiple times on the residual velocity fields that we derive by varying each time the inclination of the corresponding model velocity field in steps of 1 deg. From these harmonic decompositions, we derive the function $c_3(i)$ to determine the uncertainty on inclination. 

\begin{figure}
 \includegraphics[width=90mm]{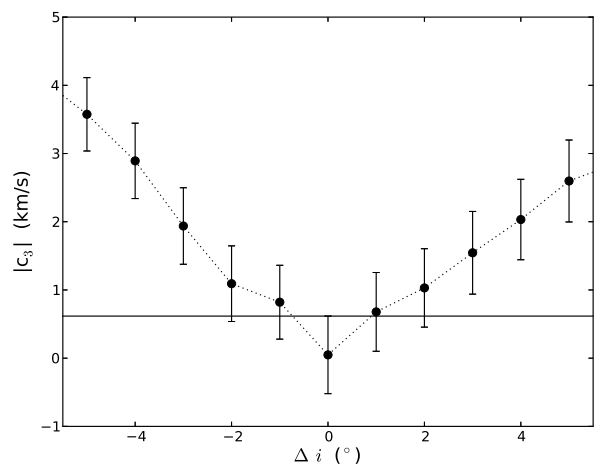}
\caption{NGC 3941 inclination angle offset versus $c_3$-amplitude. The horizontal line describes the $c_3\mathrm{(best fit)}+\Delta c_3\mathrm{(best fit)}$, where $\Delta c_3$ is the 1-$\sigma$ uncertainty of the $c_3$-term. Since the minimum occurs at the best-fitting inclination angle, and any offset from the best-fitting inclination gives rise to a $c_3$ amplitude, the inclination can be confidently constrained.}
\label{n3941_c3}
\end{figure}

We illustrate this technique showing its application to the velocity field of NGC~3941. This galaxy hosts an \hi\ ring which is well resolved, as shown in Fig. \ref{n3941nhi}. 
The best-fitting parameters resulting from a tilted-ring fit of the outer rings are $i=57$ deg and $v_{\mathrm{circ}}=148~\mathrm{km}~\mathrm{s}^{-1}$.
Figure  \ref{n3941_c3} shows the value of $c_3$ as a function of inclination offset from the best fitting inclination of 57 deg and has, of course, its minimum at the best-fitting inclination. The task \texttt{reswri} returns an error bar $\sigma_3$ on $c_3$. We define the uncertainty on $i$ as the interval $\Delta i$ within which $c_3$ is equal to its minimum value within one $\sigma_3$. The horizontal line in Fig. \ref{n3941_c3} represents this definition and, in this case, returns an error on $i$ of $\pm 1$ deg.

We can now use this result to estimate the uncertainty on the circular velocity. In this case, running \texttt{rotcur} again at fixed inclinations $56$ deg and $58$ deg respectively, and solving for the circular  velocity, yields a change of $\pm 3$ km s$^{-1}$ on the latter. In fact, in cases like this we prefer to assign a conservative estimate of $\pm 8$ km s$^{-1}$ to the $v_{\mathrm{circ}}$ uncertainty, corresponding to about half the velocity resolution.

\begin{figure*}
\hfill
\includegraphics[width=84mm]{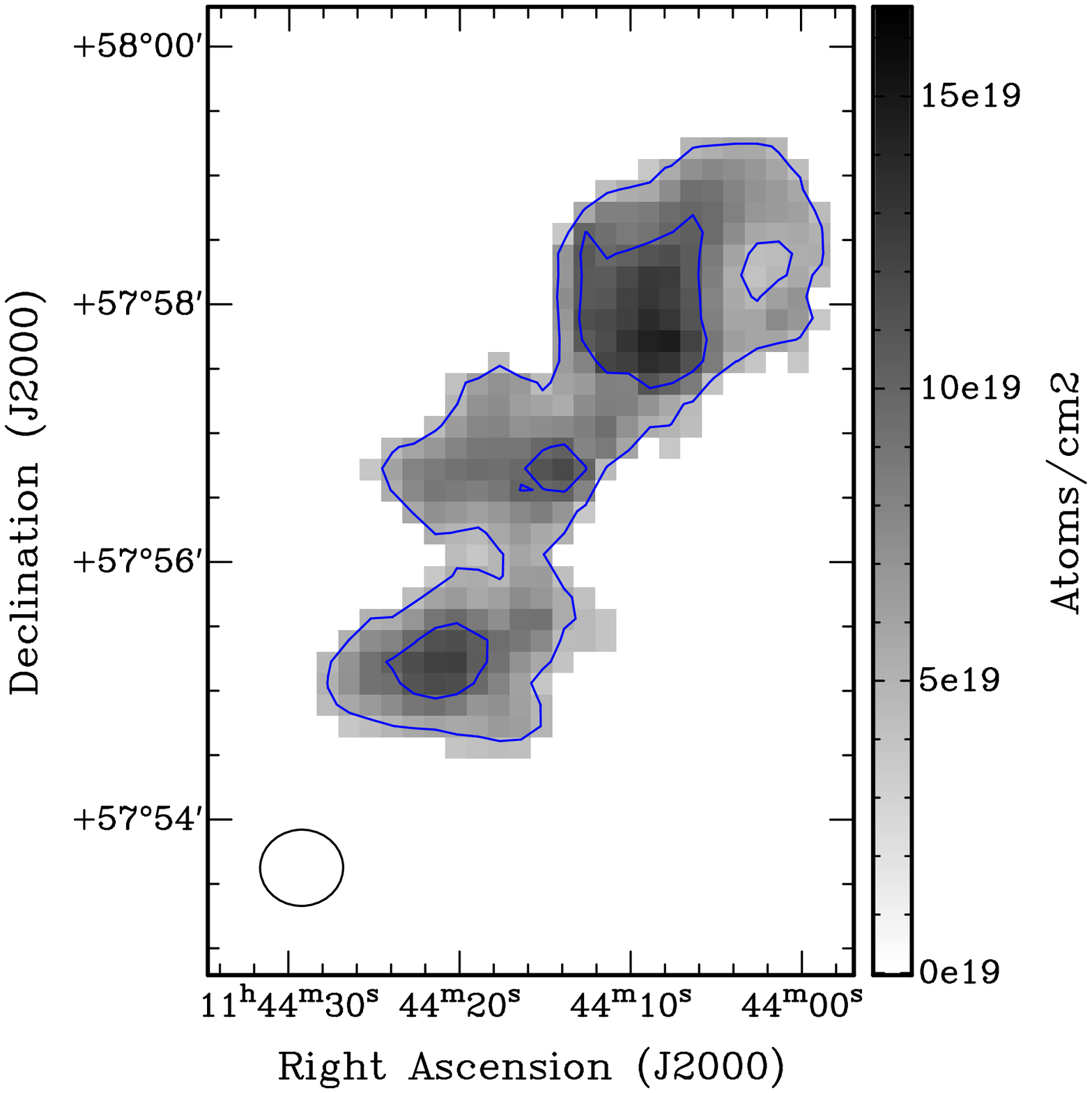}
\hfill 
\includegraphics[width=84mm]{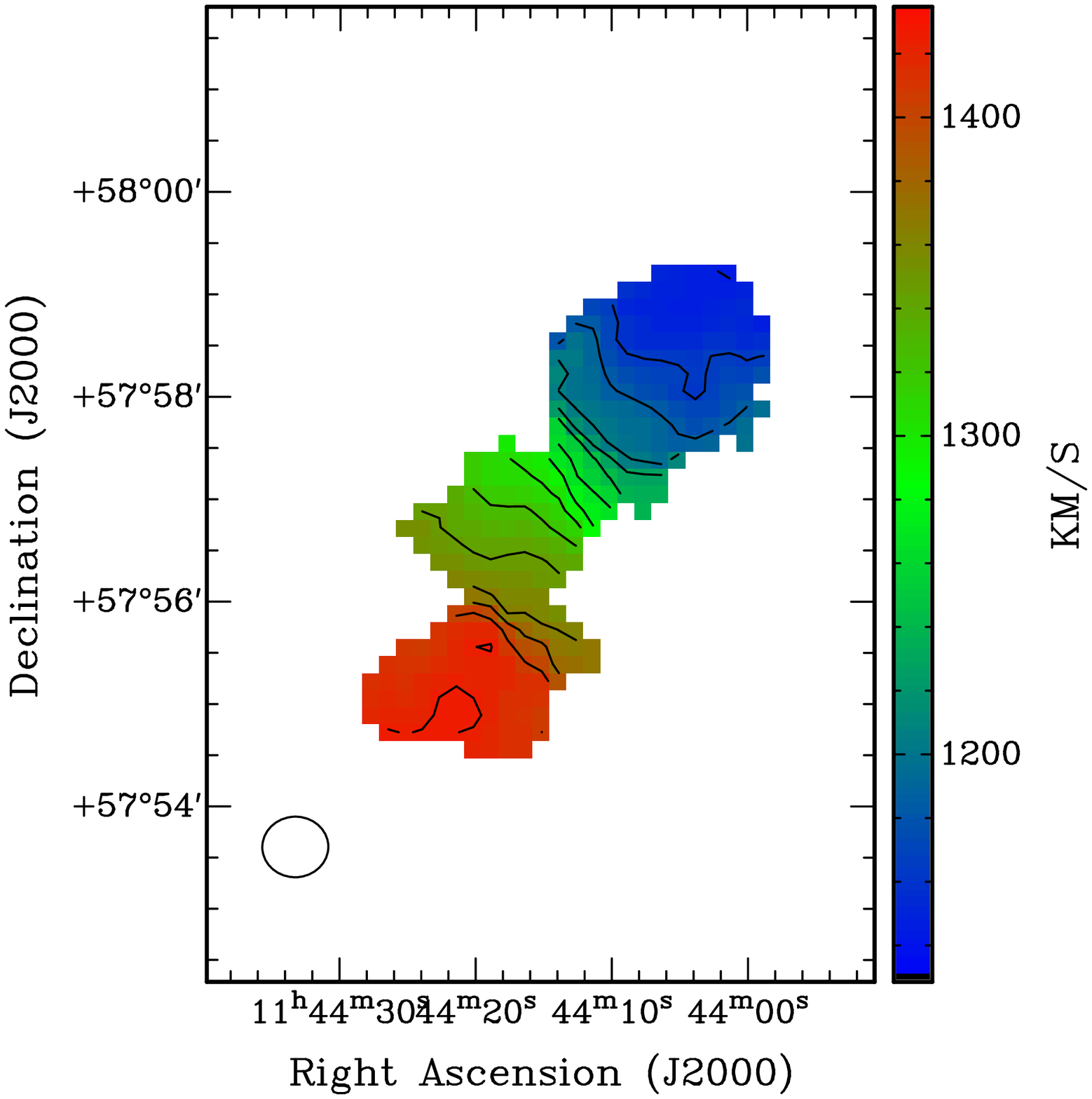}
\hfill
\caption{\textit{Left}: NGC 3838 column density map. The contours are $5, ~10 \times 10^{19}~\mathrm{cm}^{-2}$. The ellipse on the bottom-right describes the the beam shape. \textit{Right}: First-moment velocity field. The contours range from 1100 to 1450 km/s with 20 km/s steps. The ellipse on the bottom-left describes the  beam shape.} 
\label{3838_nhi}
\end{figure*}

\subsection{Model data cubes}
\label{datacubes}

For poorly-resolved galaxies, or galaxies with a significantly warped disk, beam-smearing persists even in the outer regions and we model the full \hi\ data cube instead of the velocity field (NGC 2824, NGC 3838, NGC 4203). In these cases we therefore perform a tilted-ring analysis using the \texttt{TiRiFiC} software (\citealt{jozsa2007})\footnote{http://gigjozsa.github.io/tirific/} which is specifically designed to take beam smearing effects into account.
Our goal is not to construct a perfectly matching kinematical model of the entire galaxy, but rather to construct simple models to constrain the inclination and circular velocity of the galaxy for the outermost points of the rotation curve. Since the \texttt{TiRiFiC} opitimiser is sensitive to  local $\chi^ 2$  minima, accurate initial guess values are required for the parameters. The starting values for the fit are chosen analogous to Section \ref{velfield}. After setting the initial parameters, we  fix some parameters and fitting only a few parameters at a time. This procedure is done iteratively until a stable solution emerges. For simplicity, and considering the limited scope of our fitting (i.e., focus on the outermost point of the rotation curve), we fit position angle and inclination as being constant with radius if no strong warp is present. In the case of NGC 4203, however, a warp is present and this warp is modelled as a linear function in inclination and position angle. 

\begin{figure}
 \includegraphics[width=85mm]{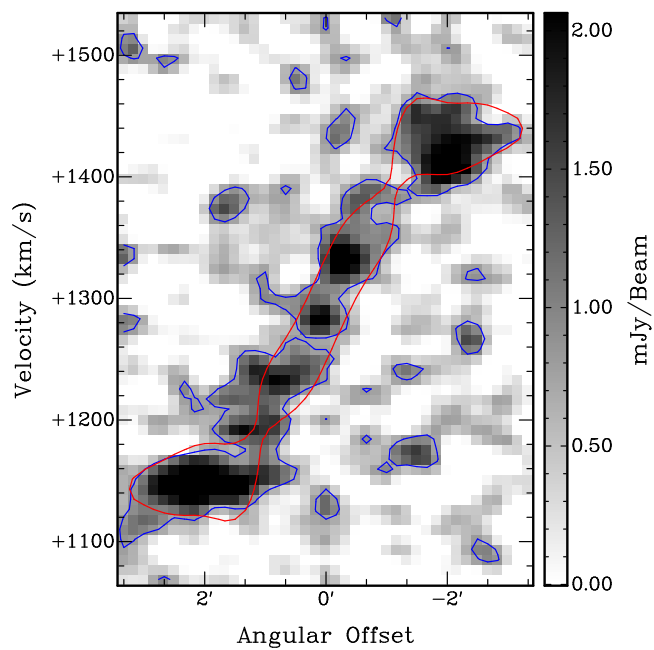}
\caption{NGC 3838 position-velocity diagram. The blue contour corresponds to the observation and the red contour to the model. In both cases, the countour level is at $0.8 ~ \mathrm{mJy}$ beam$^{-1}$, which is approximately the 2-$\sigma$ noise level of the data cube.}
\label{n3838_pv}
\end{figure}

\begin{table*}
\centering
\caption{The data relevant for the TFRs discussed in this paper.}
\begin{tabular}{ c c c c c c c c c c c }         
\hline\hline
ID 	&$D$ & Method & $i$ & $\Delta i$ &$v_{\mathrm{circ}}$	&$\Delta v_{\mathrm{circ}}$&$\log{M}_{{\rm HI}}$ &$\log L_{\mathrm{r}}$ &  $\log{M}/L_{\mathrm{r}}^{\rm JAM} $&$\log{M}/L_{\mathrm{r}}^{\rm SFH}$ \\
 	& (Mpc)& & (deg.) & (deg.) &($\mathrm{km}\,\mathrm{s}^{-1}$)&($\mathrm{km}\,\mathrm{s}^{-1}$) & $({M}_{\odot})$ & $L_{\odot}$  & ${M}_{\odot}/L_{\odot r} $ & ${M}_{\odot}/L_{\odot r} $	\\ 
(1) & (2) & (3) & (4) & (5) & (6) & (7) & (8) & (9) & (10)  & (11) \\ 
\hline
NGC 2685     &16.7	& c    &65	&4       &144	&10	    &9.33 &9.857   &0.455 	& 0.620    \\
NGC 2824     &40.7	& c    &65	&8       &162	&10     &7.59 &9.889   &0.628 	& 0.343    \\
NGC 2859     &27	  & v    &35	&5       &215	&41     &8.46 &10.404  &0.568 	& 0.732    \\
NGC 2974     &20.9	& v    &60	&2       &310	&10     &8.74 &10.152  &0.981 	& 0.787    \\
NGC 3522     &25.5	& v    &41	&2       &121	&8      &8.47 &9.600   &0.705 	& 0.648    \\
NGC 3626     &19.5	& v    &54	&1       &169	&8      &8.94 &10.102  &0.437 	& 0.261    \\
NGC 3838     &23.5	& c    &66	&8       &159	&14     &8.38 &9.772   &0.589 	& 0.694    \\
NGC 3941     &11.9	& v    &57	&1       &148	&8      &8.73 &9.940   &0.400 	& 0.693    \\
NGC 3945     &23.2	& v    &57	&--5,+1	 &237	&13     &8.85 &10.394  &0.628 	& 0.760    \\
NGC 3998     &13.7	& v    &66	&2       &246	&20     &8.45 &9.967   &0.971 	& 0.818    \\
NGC 4203     &14.7	& c    &30	&3       &197	&35     &9.15 &10.067  &0.537 	& 0.827    \\
NGC 4262     &15.4	& v    &67	&--8,+1  &198	&10     &8.69 &9.726   &0.753 	& 0.782    \\
NGC 4278     &15.6	& v    &45	&--2,+5  &256	&26     &8.80 &10.247  &0.829 	& 0.846    \\
NGC 5582     &27.7	& v    &50	&2       &258	&10     &9.65 &10.140  &0.722 	& 0.752    \\
NGC 6798     &37.5	& v    &52	&2       &190	&8      &9.38 &10.028  &0.660 	& 0.684    \\
UGC 6176     &40.1  & v    &47	&--2,+6 &144	&14     &9.02 &9.751   &0.685 	& 0.328    \\ 
\hline       
\label{tfr_data}
\end{tabular}
\tablefoot{
List of the galaxies where the circular velocities at large radii  are derived  from the \hi\ kinematics.  Column (1): the name is the galaxy designation.  Column (2): galaxy distance as listed in \citet{cappellari2011}. Column (3): c indicates that the circular velocity is based on an analysis of the full data cube, whereas v indicates a velocity-field based circular velocity. Column (4) and (5): inclination angle and uncertainty. Column (6) and (7): circular velocity and uncertainty. Column (8): \hi-mass (from \citet{serra2012}). Column (9): $r$-band luminosity (from \citealt{cappellari2013b}). Column (10): $r$-band  mass-to-light ratios from stellar dynamics (from \citealt{cappellari2013b}) and Column (11): $r$-band mass-to-light ratios from star formation histories (from Cappellari et al.\ 2013a).}
\end{table*}

Although the velocity fields do not allow for a direct derivation of the tilted-ring parameters for poorly resolved galaxies, we can still use them to estimate the uncertainties on the inclination angle derived from the data cube and therefore the uncertainty on the circular velocity. In practice, the uncertainty on $v_{\mathrm{circ}}$ is constrained by deriving a set of model velocity fields using the \texttt{GIPSY}-task \texttt{velfi} with inclinations offset from the best-fitting inclination angle. By inspecting the iso-velocity contours of these velocity fields, we estimate a confidence interval for the inclination and from this we estimate the uncertainty on $v_{\mathrm{circ}}$.

We illustrate this method showing its application to NGC 3838.
The \hi\ in NGC 3838 is faint and not well resolved by the WSRT beam, causing a large uncertainty in the intensity-weighted velocity. Fig. \ref{3838_nhi} shows the column density map. The ring as reported by \citet{serra2012} is not resolved along the minor axis and therefore a basic \texttt{rotcur} tilted-ring fit and a harmonic decomposition of the velocity field will not yield any reliable results. 
Using fits to the data cube,  we find a best fitting inclination and circular velocity of $66$ deg and 159 km s$^{-1}$, respectively. We derive the uncertainty on the latter value by building a number of \hi\ velocity fields for a range of inclination values around the best-fitting one (changing inclination in steps of $1$ deg). We estimate by eye the acceptable range of inclination, which in this case is $\pm 8$ deg. Finally, we again generate a model data cube having fixed the inclination at the boundaries of this interval and obtain an uncertainty on the circular velocity of $\pm 16$ km s$^{-1}$. Figure \ref{n3838_pv} shows a comparison of the observed position-velocity diagram with the one based on the model.

\begin{figure}
 \includegraphics[width=9cm]{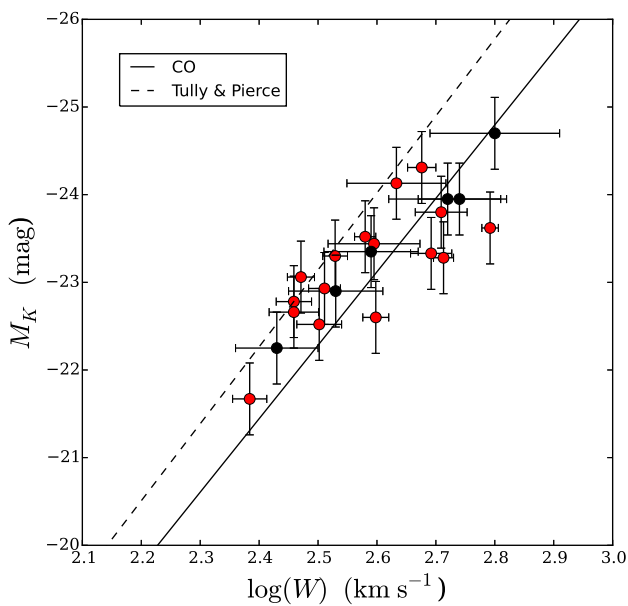}
\caption{The \hi\ Tully-Fisher relation for our 16 galaxies (red circles; $W=2v_\mathrm{circ}$). The black symbols give the data based on planetary nebulae  from \citet{Cortesi2013} for comparison. The CO TFR of \citet{davis2011} is given by the drawn line. The dashed line gives the  \citet{tully2000} $K$-band TFR for spiral galaxies. }
\label{tfr}
\end{figure}

\begin{table*}
\centering
\caption{Parameters of the \hi\ K-band and baryonic TFR.}
\begin{tabular}{c c c c c}
\hline
 TFR & slope $a$ & intercept  $b$ &  $\sigma$&  $\rho$ \\ 
\hline\hline
$K$-band \\
unconstrained & $-4.67 \pm 0.99 $ & $-23.25 \pm 0.12 $ & 0.43 & 0.74 \\
slope  fixed & $-8.38$ (CO) & $-23.31 \pm 0.17$ &0.67 &  \\
slope  fixed & $-8.78$ (TP00) & $-23.32 \pm 0.18$ & 0.69 &  \\  
slope  fixed & $-9.13$ (WBC10) & $-23.32 \pm 0.19$ &0.72 &  \\
slope  fixed & $-9.64$ (NV07) & $-23.33 \pm 0.20$ & 0.77 &  \\ 
\hline
Baryonic\\
unconstrained (JAM) &  $2.38 \pm 0.25$ & $10.71 \pm 0.03$ & 0.24 & 0.94 \\
unconstrained (SFH) & $2.51 \pm 0.42$ & $10.71 \pm 0.05$ & 0.47 & 0.80 \\
slope  fixed  (JAM) & 3.36 (NV07) & $10.72 \pm 0.04$ & 0.39 &  \\
slope  fixed  (SFH) & 3.36 (NV07) & $10.72 \pm 0.06$ & 0.57 &  \\ 
slope  fixed   (JAM) & 3.84 (McG12) & $10.73 \pm 0.05$ & 0.50 &  \\
\hline
\end{tabular} 
%\vspace*{1.5cm}
\tablefoot{
The rows in the table present the best-fitting parameters  for the $K$-band TFR (all inverse fits) and the JAM  baryonic TFR for the  16 galaxies in Table \ref{tfr_data}. The slopes without uncertainties correspond to the cases where the slope has been fixed to a reference TFR: CO: \citet{davis2011}, TP00: \citet{tully2000}, WBC10: \citet{williams2010}, NV07: \citet{noordermeer2007} and McG: \citet{mcgaugh2012}. The scatter $\sigma$ in mass of the baryonic TFR has been converted into magnitudes by multiplying by 2.5. The last column gives the correlation coefficient of the data.}
\label{table_fitresults}
\end{table*}

\section{The H\,{\small I} {\sl K}-band Tully-Fisher relation of early-type galaxies}
\label{kbandtfr}

In this section we discuss the $K$-band TFR of ETGs, compare it to the one derived using CO kinematics by \citet{davis2011} and compare our results to the TFR of spiral galaxies. The relevant data used are summarised in Table \ref{tfr_data}. 

Figure \ref{tfr} shows the \hi\ TFR as found from our data. For comparison, the figure includes the early-type galaxies whose circular  velocity is estimated using planetary nebulae by \citet{Cortesi2013}. Although we do not use these  galaxies in the formal TFR fits below, the figure shows that the two samples are fully consistent.

We determine the parameters of the TFR by fitting a relation of the form
\begin{equation}
M_{{K}}=a\left(\log W-2.6\right)+b
\end{equation}
\label{tfr_relation}
to our data. We perform such fits a number of times with different constraints using a weighted least-square fit of the inverse relation using the \texttt{MPFITEXY} package of \citet{williams2010}, which depends on the \texttt{MPFIT}-package (\citealt{markwardt2009}). We list the results in Table \ref{table_fitresults}.  Table \ref{table_fitresults} shows that the scatter around the unconstrained fit is only slightly larger than expected from errors in the data (mainly distance errors). We also find that the slope $a$ and intercept $b$ are not well constrained by such a fit. This is because the correlation $\rho$ between the points is only modest - the un-weighted correlation coefficient is 0.74. Therefore, to be able to make a better comparison with other published TFRs, we have also performed fits where we have constrained the slope  to  that of other published TFRs.

\subsection{Comparison to the CO $K$-band TFR}
\label{cokbandtfr}

We start by comparing the \hi\ K-band TFR of ETGs to the analogous CO relation from \citet{davis2011}. This may be interesting because, since the    typical CO disk extends only to about 1 $R_{\rm eff}$ while our \hi\ data extends much further, it allows to compare the TFR valid for the inner regions of ETGs with the one of the entire galaxy. This, in turn, may tell us something about the shape of the mass distribution. In \citet{davis2011},   a number of TFR relations are derived using different samples and methods. Here we compare  our data to their ``hybrid' TFR, which is based on the largest sample of galaxies for which either single-dish or interferometric CO data suggest that the molecular gas has reached the flat part of the rotation curve.

We investigate a potential offset between \hi\ and CO TFR by keeping the slope $a$ fixed to the CO value of $-8.38$. The resulting intercept is $b = -23.31 \pm 0.17$ (Table \ref{table_fitresults}), to be compared with the CO value of $-23.12 \pm 0.09$. Therefore, the difference between CO- and \hi\ TFR intercept is constrained to be between $+0.45$ and $-0.07$. At fixed $M_{K}$, this corresponds to a ratio of \hi\ to CO circular velocity between 0.88 and 1.02. Therefore, the circular velocity of a typical ETG decreases between 0 and 10 percent when going from 0.5 - 1 $R_{\rm eff}$, where it is traced by CO, to the outer regions, where it is traced by \hi. \citet{cappellari2015} recently found that the mass profiles of a sample of 14 ETGs can be well described by a nearly isothermal power law $\rho_{\rm tot} \propto r^{-\gamma}$ with an average logarithmic slope $\gamma = 2.19$ with remarkable small scatter in $\gamma$. The small decrease in circular velocity we observe is consistent with such mass profiles. We will explore the mass profiles of our sample galaxies, combining the \atlas\ data on the stellar kinematics with our \hi\ data,  in more detail in a future paper (Serra et al.\ in prep.).

\subsection{Comparison to the $K$-band TFR for spirals}
\label{spiraltfr}

Another interesting exercise is to compare our data with the TFR found for later-type galaxies.
\citet{davis2011}  reported an offset between their CO TFR for ETGs of the \atlas\ sample with the one for spiral galaxies in the sense that for a given circular velocity, ETGs are fainter. Our data also indicate such an offset (Fig.\ \ref{tfr}).  However, since our \hi\ TFR for early-type galaxies is slightly offset from the one based on  CO data in the other direction, the offset from the spiral relation is somewhat smaller. If we fit a relation with the slope fixed to that of the TFR found for spiral galaxies by \citet{tully2000}, we find an offset from their relation of 0.70 magnitudes, compared to 0.93 mag found found by \citet{davis2011}. We therefore find a shift between the relations of about 0.2 mag.   

We can also compare our results with those of \citet{williams2010}, who studied an ensemble of 14 S0 and 14 spiral galaxies and found their S0 TFR to be  0.5 mag below their spiral TFR. A fit to our data fixing the slope to that found by \citet{williams2010}  gives a 0.25 mag offset from their sample of spirals.

\begin{figure}
 \includegraphics[width=9cm]{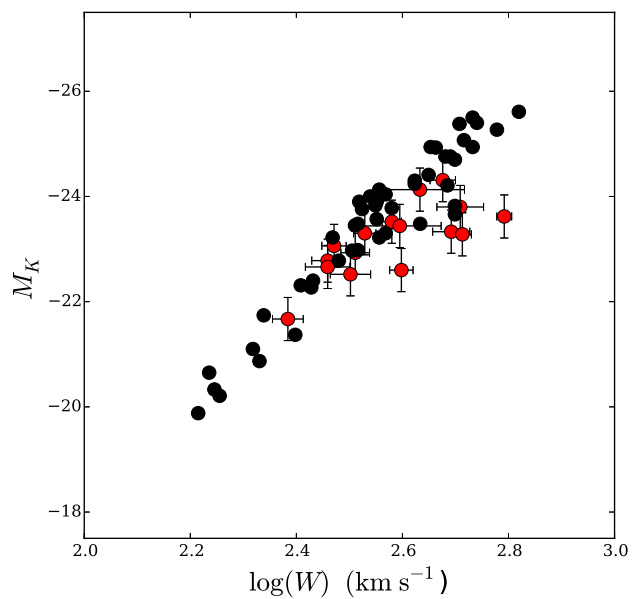}
\caption{Our data (red symbols) compared with the TFR derived by \citet{noordermeer2007} using the asymptotic circular velocities (black symbols).}
\label{nv07_ktfr}
\end{figure}

Finally, 
\citet{noordermeer2007} study the $K$-band TFR of massive spirals. Noordermeer \& Verheijen find a $K$-band TFR of the form $M_K = (-9.64 \pm 0.25)(\log W -2.6)+(-23.95 \pm 0.04)$. A fit to our data with the slope fixed to the slope found by \citet{noordermeer2007}  (see Table \ref{table_fitresults}) shows an offset of  0.62 mag.

Our analysis appears to confirm a zero-point offset of the $K$-band ETG TFR with respect to the TFR for spirals of about 0.5-0.7 mag. \citet{davis2011} suggested that a simple fading mechanism (see e.g.\  \citealt{1980ApJ...236..351D}) combined with low levels of residual star formation (as implied by the presence of molecular gas) would explain such an offset. \citet{williams2010} remark that a difference in size between ETGs and spirals could provide an alternative explanations for the offset. Their work showed that an offset between their S0 TFR and spiral TFR is also present in the dynamical mass TFR. Dynamical mass scales as $r v_{\mathrm{circ}}^2$, where $r$ is some characteristic radius.  Therefore, if ETGs were more concentrated than spirals, an offset would occur between the relations for these two types of galaxies. Below we show that no significant offset is present in the baryonic TFR, suggesting that the differences in mass distribution do not play an important role.

\begin{figure}
 \includegraphics[width=9cm]{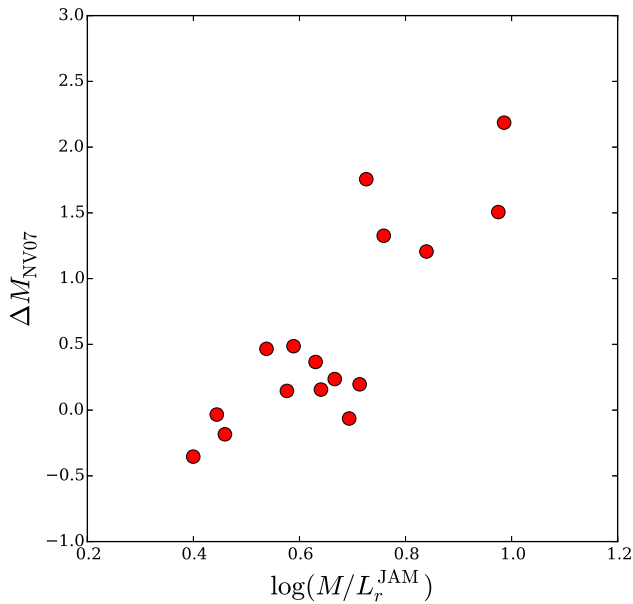}
\caption{Residual (in magnitude) of our galaxies from the TFR published by \citet{noordermeer2007} as function of mass-to-light ratio ($r$ band) derived by \citet{cappellari2013b} using dynamical modelling.}
\label{off07_ktfr}
\end{figure}

In interpreting these numbers, one should take into account that the various studies use velocities measured in different ways and for different regions of the galaxies. For example, the \citet{williams2010} analysis refers to the central regions while that of \citet{noordermeer2007} uses \hi\ circular velocities at large radius and is more similar to ours. In this respect, it is interesting to note that \citet{noordermeer2007} found that, depending on which velocity measure they used, there is a subset of massive spiral galaxies rotating 'too fast' (or equivalently, that are 'too dim'), causing a 'kink' or a change of slope in the relation. This effect is smallest for their sample if they use the circular velocity derived for large radius as opposed to one based on the maximum velocity more representative for the inner regions. They interpret this as an indication that the shape of the rotation curve changes systematically with mass in the sense that more massive galaxies tend to have a rotation curve that is slightly declining while it is more flat for less massive spirals. Above we found evidence that the rotation curves of our early-type galaxies are behaving similarly to the massive spiral galaxies of  \citet{noordermeer2007}.

\begin{figure}
\includegraphics[width=9cm]{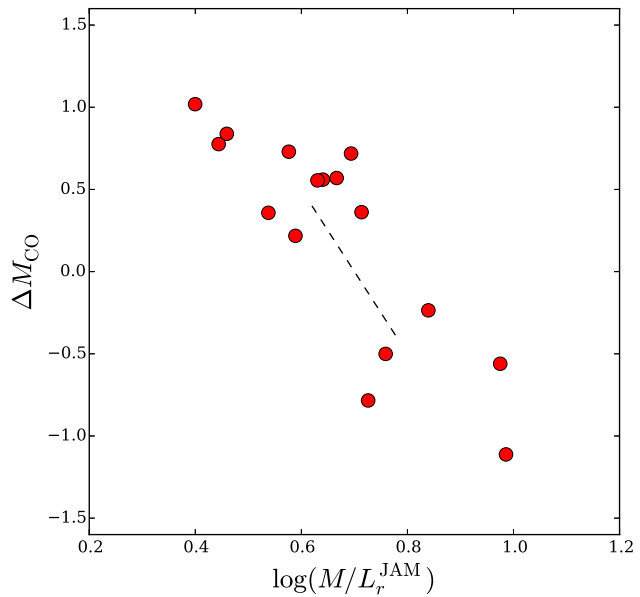}
\caption{Residual magnitudes with respect to the CO TFR from \citet{davis2011} as function  of  ${M}/L_{\mathrm{JAM}}$. The dashed line indicates how a 20\% distance error  moves an object in the plot. Such an error corresponds  to an error in  absolute magnitude of about 0.4 mag and of 0.07 in $\log {M}/L_{\mathrm{JAM}}$}
\label{ml-cores}
\end{figure}

Figure \ref{nv07_ktfr} shows the \citet{noordermeer2007} data for spiral galaxies, based on the asymptotic velocity, compared to our data for the \atlas\ galaxies.  Our galaxies still show an offset to fainter magnitudes despite the fact that we also use the circular velocity at large radius so the effects of decreasing circular velocities should be largely  taken out.  This suggests that differences in stellar population are the main cause for the observed offsets.

To further investigate this, we note that Fig.\ \ref{nv07_ktfr}, as well as the results given in Table 2, show that the slope of the TFR for ETGs is flatter than that of later-type galaxies, pointing to some systematic effect. Figure \ref{off07_ktfr} shows the offset of our galaxies from the TFR of \citet{noordermeer2007} as a function of the $r$-band mass-to-light ratio derived by \citet{cappellari2013b} using dynamical modelling of the observed stellar kinematics (see below).  There is a clear trend visible in this figure, which  underlines that differences in stellar populations play a major role when comparing the TFRs of early-type and late-type galaxies and that a more accurate scaling relation, based on mass instead of light, exists.

\begin{figure*}
\hfill
\includegraphics[width=8.5cm]{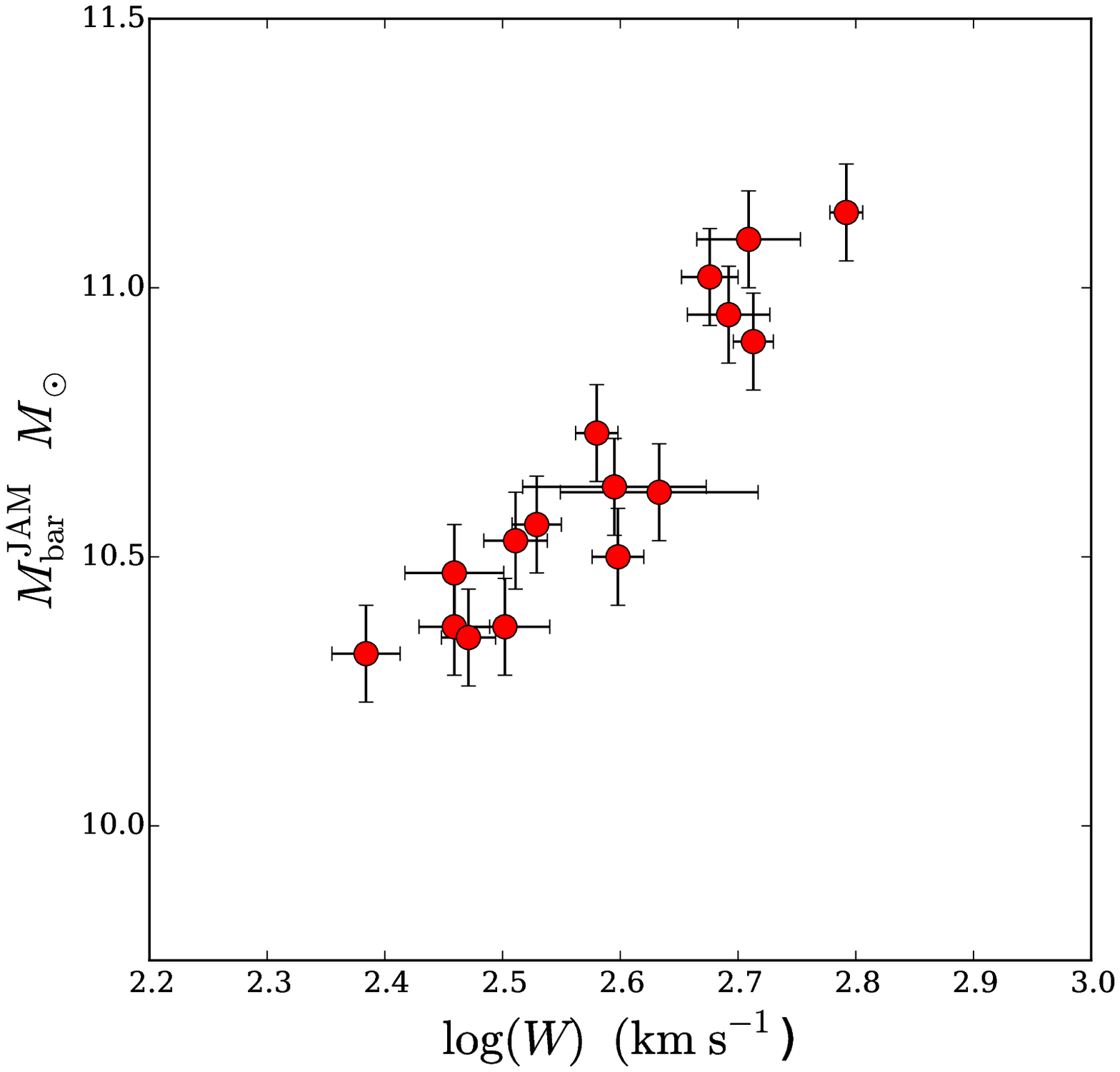}
\hfill
\includegraphics[width=8.5cm]{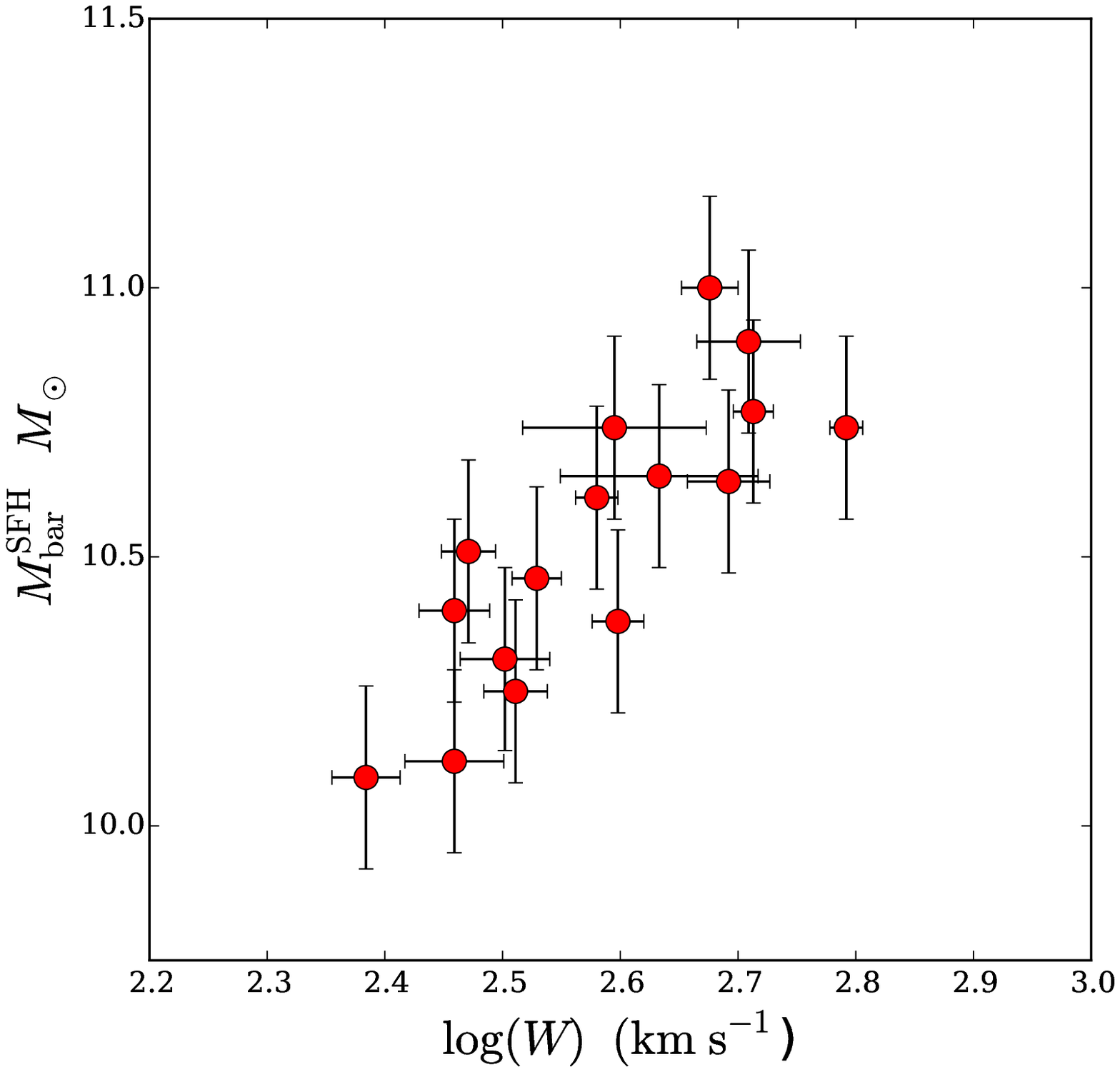}
\hfill
\caption{The baryonic TFR derived using dynamically determined $M/L$'s (left) and  $M/L$'s derived from modelling the star formation history (right). }
\label{btfr_jamsfh}
\end{figure*}

This is also suggested by Fig.\  \ref{ml-cores} where we plot the residual (in magnitude) of our galaxies from the CO TFR as function of the $M/L$ as in Fig.\  \ref{ml-cores}. It is clear that  also  here a systematic trend is present, albeit with a different slope than in Fig.\ \ref{off07_ktfr} due to the fact that the CO TFR of \citet{davis2011} has a flatter slope than the TFR given by \citet{noordermeer2007}. The fact that a correlation exists between the offset from the TFR and the $M/L$ of the stellar population suggests that the scaling relation between baryonic mass and circular velocity will have smaller scatter than the one involving light and circular velocity.

Before we investigate this, one should note, however, that both the absolute magnitude as well as ${M}/L_{\mathrm{JAM}}$ are distant dependent quantities and therefore both suffer from distance errors which could, in principle, cause a correlation of the kind seen in Fig.\ \ref{ml-cores}. In fact, the absolute magnitude depends on $D^2$ while ${M}/L_{\mathrm{JAM}}$ depends on $D^{-1}$.  In Fig.\  \ref{ml-cores} is indicated, with the dashed line, how distance errors would move data points through the figure. Figure \ref{ml-cores} shows that the data points follow a  similar slope as would be expected from distance errors, but that the observed spread is much larger than expected from the errors in distance for our sample galaxies. If  the observed spread would be entirely due to distance errors, it would imply that the distances are uncertain up to a factor 2 which is much larger than the typical $\sim$20\% distance error estimated in Sec.\ \ref{sample} (see also \citealt{cappellari2011}).

\section{The baryonic TFR  of early-type galaxies}
\label{btfr_section}

The TFR is an empirical relation between circular velocity and luminosity which may result from a more fundamental relation between  galaxy properties. In this context,  a number of  authors have investigated the existence of a baryonic TFR, where total baryonic mass replaces luminosity (see e.g.\  \citealt{mcgaugh2000,mcgaugh2012}). It has been suggested that the baryonic TFR  is a more  fundamental relation since it appears to have less scatter than the classical TFR (see e.g.\  \citealt{2004PASA...21..412G, 2008MNRAS.386..138B}, \citealt{noordermeer2007}, \citealt{zaritsky2014}). Almost all earlier studies of the baryonic TFR have focussed on spiral galaxies. If, as we concluded in the previous Section, systematic differences in stellar population are important for explaining systematic trends in the residuals of ETGs from the standard TFR, it is  interesting to include early-type galaxies when considering the baryonic TFR.

We compute the baryonic mass of a galaxy as the sum of the cold-gas mass and the stellar mass, although for our sample galaxies, the stellar mass is the dominant contribution; e.g., the \hi-to-stellar mass ratio ranges between 0.1 to 8 \% (2.2 \% on average).
The atomic gas mass is computed as the \hi\ mass given in \citet{serra2012}, multiplied by a factor of 1.4 to take into account the presence of helium and metals. Moreover we add the mass of the molecular component from \citet{young2011}, which is also only a very small contribution.

\begin{figure*}
\hfill
\includegraphics[width=8.5cm]{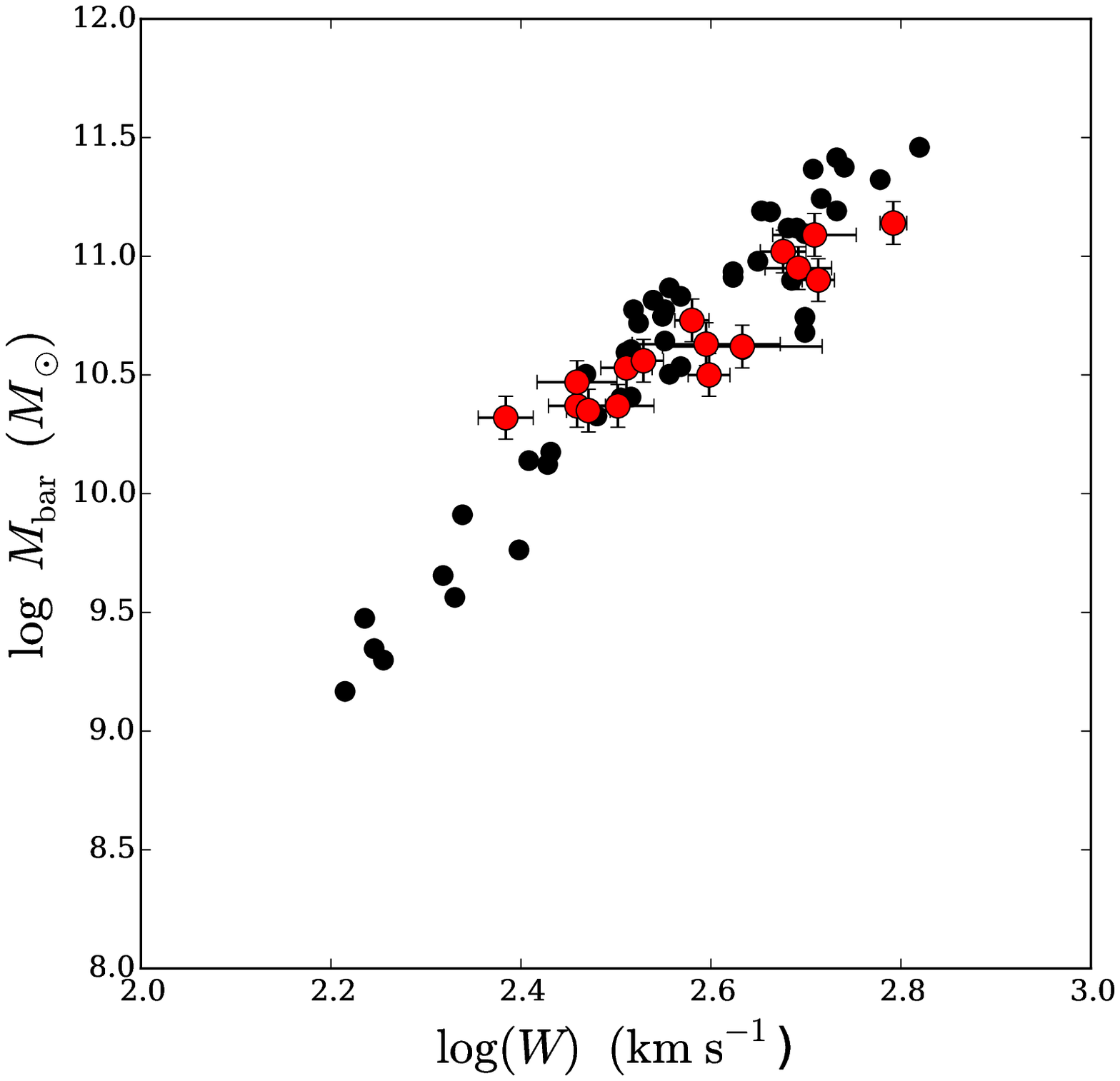}
\hfill
\includegraphics[width=8.5cm]{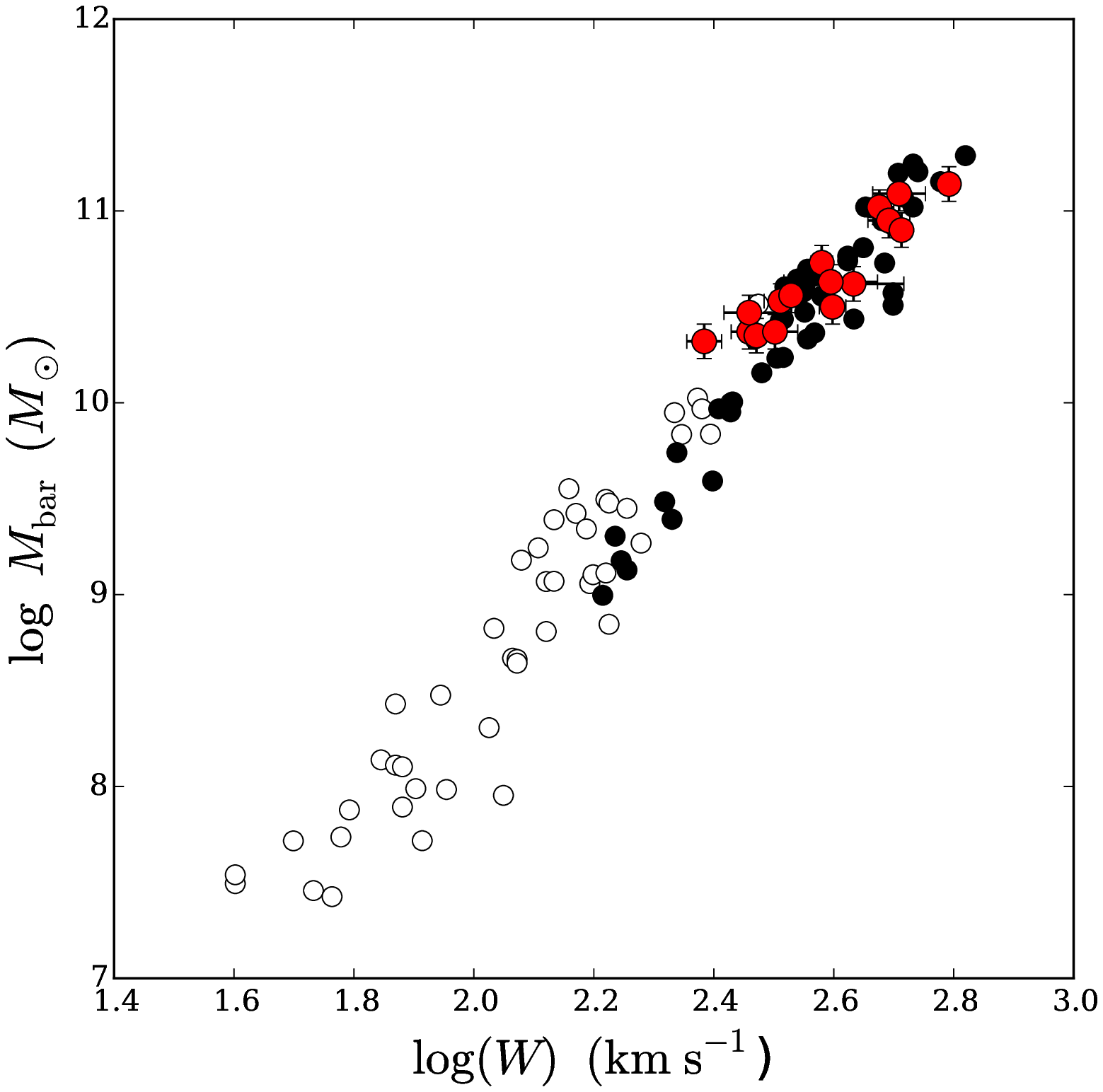}
\hfill
\caption{Baryonic TFR where the galaxy masses for our sample galaxies are estimated using ${M}/L_{\mathrm{JAM}}$. The red symbols represent our \atlas\ sample, the black symbols are the data from \citet{noordermeer2007}, while the open symbols show the data from \citet{mcgaugh2012} for gas-dominated galaxies. For the data from \citet{noordermeer2007} in the left-hand panel ${M}/L_{K}=0.8$ ${M}_{\odot}/L_{\odot}$ was used, as in the original \citet{noordermeer2007} paper, while in the right-hand panel ${M}/L_{K}=0.54$ ${M}_{\odot}/L_{\odot}$ was used. The scaling of the lefthand figure is the same as that of Fig.\ \ref{nv07_ktfr} to facilitate easy comparison.
}
\label{btfrAll}\label{btfrNV}

\end{figure*}

Stellar masses are calculated using the $r$-band luminosities from \citet{cappellari2013b} and multiplying them with a mass-to-light ratio ${M}/L_{\mathrm{r}}$. We consider two different estimates of this ratio: firstly, the mass-to-light ratios resulting from JAM modelling of  the stellar kinematics, derived for the full \atlas\ sample by \citet{cappellari2013b} where we use their results from the self-consistent JAM models (their model A). The second set of mass-to-light ratios we use are based on the star-formation histories (SFHs) derived using a Salpeter IMF (Cappellari et al.\ 2013a). Note that Cappellari et al.\ (2012), based on a comparison of these sets of mass-to-light ratios, found that there are systematic variations in the IMF among early-type galaxies and that assuming a Saltpeter IMF for all galaxies will lead to systematic effects in the estimates of the galaxy masses.

Figure \ref{btfr_jamsfh} shows the  baryonic TFRs derived for the  two different sets of mass-to-light ratios, and Table 2 shows the parameters of the best-fitting relations for 3 different sets of assumptions.. 
The baryonic relation derived using the dynamically determined ${M}/L_{\rm JAM}$ shows less scatter than the one based on the $M/L$ derived from modelling the star formation histories of the galaxies. In addition, the scatter in the SFH based baryonic TFR is very similar (when converted to optical magnitudes) to that of the standard TFR while the scatter in the JAM-based TFR is significantly smaller. Both these features are exactly what is expected if the underlying baryonic TFR is  tight while the  main sources of scatter are the uncertainties in the distances to the galaxies.  
Given that  ${M}/L_{\rm SFH}$ is  distance independent, whereas ${M}/L_{\rm JAM}$ has a $D^{-1}$ distance dependence, the errors on the  masses using ${M}/L_{\rm SFH}$ vary as $D^2$  whereas those on the  masses  using ${M}/L_{\rm JAM}$ vary only linearly with $D$. Therefore a 20\% distance error implies  a typical uncertainty of $\sim$$0.17$ dex on $\log{{M}_\mathrm{bar,SFH}}$  (or of 0.43 when converted to magnitudes, as in Table \ref{table_fitresults}) and  of $\sim$$0.09$ dex  on $\log{{M}_\mathrm{bar,JAM}}$ (or 0.23 when converted to magnitudes). The observed scatter of the baryonic TFRs of our sample is only slightly larger than these estimates. Given that the galaxy masses based on ${M}/L_{\rm JAM}$ suffer less from distance errors, and also that they suffer less from systematic errors than masses derived using   ${M}/L_{\rm SFH}$, in the following we only further consider the baryonic masses using ${M}/L_{\rm JAM}$.
  
Table \ref{table_fitresults} gives the results of fits, of the form 
\[
\log {M_{\rm bar}} = a\ (\log W - 2.6) + b,
\]  
to our data for a number of cases. This table shows that, if both the slope $a$ and the intercept $b$ both are free parameters, the slope is flatter than that of the baryonic TFR for spiral galaxies derived by \citet{noordermeer2007} and \citet{mcgaugh2012}, although, given the small sample size and the limited range in log $W$ covered by our sample, this result is somewhat uncertain.

In Fig.\ \ref{btfrNV} we compare the baryonic TFR with that derived by \citet{noordermeer2007} for their sample of massive spiral galaxies.  In deriving baryonic masses, they assumed, in a quite different approach to what we have chosen, a fixed ${M}/L_{K}=0.8$ ${M}_{\odot}/L_{\odot}$ for all galaxies in their sample, a value  motivated by  the maximum-disk scenario. Despite the different approaches, a comparison of Fig.\ \ref{btfrNV} with Fig.\ \ref{nv07_ktfr} shows that our early-type galaxies lie much closer to the spiral baryonic TFR of \citet{noordermeer2007} than they do to the standard TFR of \citet{noordermeer2007}.   When we fit our baryonic TFRs with the same slope as found by \citet{noordermeer2007} for their relation,  we find an intercept of $10.71 \pm 0.03$ compared to $10.88 \pm 0.02$ for \citet{noordermeer2007} sample. This implies an offset between the baryonic TFR of spirals and of ETGs of about $0.17 \pm 0.04$ dex in mass, or when expressed as magnitudes, it is 0.43 mag (compared to 0.62 mag for the standard TFR). 
Recent work has suggested that the disks of spiral galaxies are not maximal \citep[e.g.,][]{bershady2010,martinsson2013,mcgaugh2015}  so this offset may simply reflect that using ${M}/L_{K}=0.8$ $M_\odot/L_\odot$  overestimates the stellar masses of spiral galaxies. Transforming the offset between our baryonic TFR and the one of \citet{noordermeer2007} into a reduction of ${M}/L_{K}$ suggests that one should lower this quantity with a factor 0.67 so that ${M}/L_{K}=0.54$ $M_\odot/L_\odot$ would be a more appropriate value for spiral galaxies, in line with the results of some recent work \citep[e.g.][]{bershady2010,martinsson2013,mcgaugh2015}. 

In the right-hand panel of Fig.\ \ref{btfrNV} we  explore how our results compare to the baryonic TFR for low-mass, gas-rich galaxies. We compare our data with the baryonic TFR presented by \citet{mcgaugh2012}. This latter sample consists of particularly gas-rich galaxies (${M}_{\mathrm{gas}} > {M}_{\rm star}$) to minimise the influence of errors in the assumed the stellar mass-to-light ratios which have been estimated by \citet{mcgaugh2012} using a Kroupa initial mass function. In the right-hand panel of Fig.\ \ref{btfrNV} we also include the data from \citet{noordermeer2007} where we now assumed a fixed ${M}/L_{K}=0.54$ ${M}_{\odot}/L_{\odot}$ for all their galaxies. Figure \ref{btfrNV} shows that, despite  the different methods used for computing stellar masses, the combined dataset produces a well-defined baryonic TFR that seems to be valid for small gas-rich galaxies, massive spiral galaxies as well as early-type galaxies. \citet{mcgaugh2012}  fitted a relation of the form $\log {M_{\rm bar}} = 3.82\ (\log W - 2.6) + 10.79$.  When we fit a baryonic TFR with the same slope to our data (Table \ref{table_fitresults}) we find a small, but not very significant offset, of $0.06 \pm 0.06$  dex for our galaxies.
This is quite different from the situation for the standard TFR where the early-type galaxies lie clearly offset from the spiral TFR. The well-defined baryonic TFR shows that the offsets from the standard TFR are due to different stellar populations between the two classes of galaxies and that the fundamental relation is more likely to be the baryonic TFR.

\section{Conclusions}

We study the \hi\ $K$-band TFR and the baryonic TFR of a sample of 16 ETGs taken from the \atlas\ sample which all have very  regular \hi\ disks that extend  well beyond the optical body ($\gtrsim 5R_{\rm eff}$). We use the kinematics of these disks to estimate the circular velocity at large radius for these galaxies. For galaxies that are sufficiently well resolved spatially, we use the velocity fields, including a harmonic analysis, to find the best-fitting $v_{\mathrm{circ}}$ for the outermost point of the rotation curve. In three other cases, we model the full data cube in order to derive the kinematical information of the galaxy, and in two additional cases, we use kinematical data from the literature. We use these circular velocities to construct the traditional $K$-band TFR as well as the baryonic TFR. In the case of the baryonic TFR, we use both  mass-to-light ratios derived from stellar kinematics and from star formation histories. We find the following:

\begin{enumerate}

\item Comparison with published TFRs derived for samples of spiral galaxies suggests that the TFR for ETGs is offset by about 0.5-0.7 mag in the sense that early-type galaxies are dimmer for a given circular velocity. The residuals from the standard TFR correlate with estimates of the $M/L$ of the stellar populations, 
suggesting that the offset is mainly driven by differences in stellar populations. 
 
\item There is a small offset between the TFR derived from circular velocities at large radius from our \hi\ data with the relation derived for the \atlas\ sample using CO data covering the galaxies' inner regions ($\lesssim 1R_{\rm eff}$). This offset suggests that the circular velocities at large radius are about 10\% lower than the CO circular velocities near 1 $R_{\rm eff}$. Such a decrease is consistent with recent determinations of the shape of the mass profiles of ETGs \citep{cappellari2015}.

\item The baryonic TFR is distinctly tighter than the standard TFR,  in particular when using the mass-to-light ratios based on dynamical models. In both cases the scatter is mainly due to distance uncertainties. From the offset we find between   our baryonic TFR with that of  \citet{noordermeer2007} derived for spiral galaxies, we find that the ETGs fall on the spiral baryonic TFR if one assumes ${M}/L_{K}=0.54$ $M_\odot/L_\odot$ for the stellar populations of the spirals. Such a value  would imply that the disks of spiral galaxies are about 60-70\% of maximal, as is also found by other recent studies of the dynamics of spiral galaxies \citep[e.g.][]{bershady2010,martinsson2013,mcgaugh2015}.
 The galaxies of our sample also fall on the baryonic TFR derived for gas-dominated, late-type, lower-mass galaxies. The fact that there is no significant offset between the baryonic TFRs of spirals and ETGs suggests that the differences in concentration of the mass distributions of these two types of galaxies have little effect on the baryonic TFR. Our analysis increases the range of galaxy morphologies for which the baryonic TFR holds, strengthening previous claims that this is a more fundamental scaling relation than the classical relation.

\end{enumerate}

\begin{acknowledgements} 

Milan den Heijer was supported for this research through a stipend from the International Max Planck Research School (IMPRS) for Astronomy and Astrophysics at the Universities of Bonn and Cologne. J\"urgen Kerp and Milan den Heijer thank the Deutsche Forschungsgemeinschaft (DFG) for support on the grants KE 757/7-2 and KE 757/9-1.
MC acknowledges support from a Royal Society University Research Fellowship.
This work was supported by the rolling grants 'Astrophysics at Oxford' PP/E001114/1
and ST/H002456/1 and visitors grants PPA/V/S/2002/00553, PP/E001564/1 and ST/H504862/1
from the UK Research Councils. TAD acknowledges the support provided by an ESO fellowship. TN acknowledges support from the DFG Cluster of Excellence 'Origin and Structure of the Universe'. AW acknowledges support of a Leverhulme Trust Early Career Fellowship. The research leading to these results has received funding from the European Community's Seventh Framework Programme (/FP7/2007-2013/) under grant agreement No. 229517. The authors acknowledge financial support from ESO.
\end{acknowledgements}

\nocite{cappellari2013a,cappellari2013b,mcgaugh2012}
\label{lastpage}

\end{document}